\documentclass[a4paper,10pt]{article}

\usepackage[numbers,sort&compress]{natbib}

\usepackage[english]{babel}

\usepackage[utf8]{inputenc}
\usepackage[T1]{fontenc}

\usepackage{mathpazo}
\usepackage{amsmath,amssymb}
\usepackage{bm} 
\usepackage{bbm} 

\usepackage[final]{graphicx}  

\newcommand{\ket}[1]{\vert#1\rangle}
\newcommand{\bra}[1]{\langle#1\vert}
\newcommand{\braket}[2]{\langle#1\vert#2\rangle}

\newcommand{\proj}[1]{\ket{#1}\!\bra{#1}}
\newcommand{\mean}[1]{\langle #1 \rangle}
\newcommand{\abs}[1]{\lvert#1\rvert}
\newcommand{\Abs}[1]{\big\lvert#1\big\rvert}
\newcommand{\norm}[1]{\lVert#1\rVert}

\DeclareMathOperator{\Tr}{Tr}

\newcommand{\re}{\operatorname{e}}

\newcommand{\beq}{\begin{equation}}
\newcommand{\eeq}{\end{equation}}

\begin{document}

\title{A critical view on transport and entanglement in models of photosynthesis}

\author{Markus Tiersch%
\thanks{Electronic address: markus.tiersch@uibk.ac.at}~
\thanks{Institute for Quantum Optics and Quantum Information, Austrian Academy of Sciences, Technikerstr.~21A, A--6020 Innsbruck, Austria;
Institute for Theoretical Physics, University of Innsbruck, Technikerstr.~25, A--6020 Innsbruck, Austria} \and 
Sandu Popescu%
\thanks{H. H. Wills Physics Laboratory, University of Bristol, Tyndall Avenue, Bristol BS8~1TL, UK} \and
Hans J. Briegel\footnotemark[2]
}
\date{}

\label{firstpage}
\maketitle


\begin{abstract}
We revisit critically the recent claims, inspired by quantum optics and quantum information, that there is entanglement in the biological pigment protein complexes, and that it is responsible for the high transport efficiency. While unexpectedly long coherence times were experimentally demonstrated, the existence of entanglement is, at the moment, a purely theoretical conjecture; it is this conjecture that we analyze.
As demonstrated by a toy model, a similar transport phenomenology can be obtained without generating entanglement. Furthermore, we also argue that even if entanglement does exist, it is purely incidental and seems to plays no essential role for the transport efficiency. We emphasize that our paper is \emph{not} a proof that entanglement does not exist in light-harvesting complexes -- this would require a knowledge of the system and its parameters well beyond the state of the art. Rather, we present a counter-example to the recent claims of entanglement, showing that the arguments, as they stand at the moment, are not sufficiently justified and hence cannot be taken as proof for the existence of entanglement, let alone of its essential role, in the excitation transport.
\end{abstract}



\section{Introduction}

In recent years, following the development of quantum information, the phenomenon of quantum entanglement has been identified as being one of the most important aspects of quantum mechanics. It was realized that the presence of entanglement confers quantum systems significantly enhanced power for accomplishing many tasks~\cite{HorodeckiRev}, such as exponentially increased speed-up of computation and significantly enhanced communication capacity. As such it has been very natural to enquire whether biological systems could have evolved to make use of entanglement. Recently, the importance of investigating this question received a major impetus following seminal experiments that indicated the existence of unexpectedly long-time coherent effects in photosynthesis~\cite{Engel2007,Collini2010,Panit2010}. Since coherence is a pre-requisite for entanglement, its discovery raises the possibility that entanglement is also present. Moreover, it is known that the energy transport in light-harvesting complexes is extremely efficient -- could it be the case that entanglement is responsible for this efficiency? A few ground-breaking studies~\cite{Caruso2009,Sarovar2010}, followed by an increasing body of literature~\cite{Ishizaki2010,Fassioli2010,Caruso2010,Bradler2010,Scholak2011,Whaley2011} raised this question and suggested that this is the case. Here we take a critical look at these results.

The main issue to be discussed here is that of coherence versus entanglement. They are definitely not one and the same thing. The existence of entanglement is a stronger criterion than the presence of coherence. That is, entanglement requires the presence coherence, but coherence does not imply entanglement in general. While coherent phenomena can also appear in systems of classical wave mechanics, entanglement is a genuine quantum phenomenon, which is required to violate a Bell inequality, for example.

It is important to mention from the beginning that while the existence of coherence in light-harvesting systems has been experimentally tested (at least in laboratory conditions), the existence of entanglement has been not. This is not surprising -- experimentally proving entanglement is a far more difficult task~\cite{GuehneRev}. As such all the discussions about entanglement in light-harvesting systems is, at present, purely theoretical.

The existence of coherence in these systems seems by now to be well established, and we are not challenging that. It is also quite reasonable to expect that coherence plays an important role in the transport problem, distinguishing it from classical diffusive processes; we are not challenging this either. What concerns us here is the existence of entanglement and its role, if it exists.
Specifically, the main questions we address in our paper are:
\begin{itemize}
\item	Are the assumptions that led to the present claims of entanglement in photosynthesis justified?
\item	Even if these assumptions were justified and entanglement would exist along the lines of those models, does the entanglement play any significant role in enhancing the efficiency of transport or it is of no consequence?
\end{itemize}
We suggest that, despite the sizable body of literature claiming the contrary, the answer to both questions is ``no''.

To be clear, we do not prove that there is no entanglement in photosynthesis. To do that would require knowledge of the system that is well beyond what is available at present.  All we do here is to present arguments that point to potential problems in the present claims that entanglement exists and plays a significant role. Our paper should rather be viewed as a counter-example and it is meant to sharpen the further investigation of the problem.

\section{Basic Issues}

Light harvesting complexes in plants and photosynthetic bacteria are comprised of protein scaffolds into which pigment molecules are embedded, e.g.\ chlorophyll or bacterio-chlorophyll molecules.
The pigment molecules absorb light in the visible or infrared part of the spectrum, and the resulting electronic excitation (exciton) is transported between the pigment molecules until it reaches a reaction center complex, where its energy is converted into separated charges.

The FMO protein complex of green sulfur bacteria is a trimeric complex that links the chlorosome antenna with the reaction center. Within each of the subunits there are seven chlorophyll molecules in close connection with each other.
Each of these molecules can be considered as a ``site'' where the excitation propagating from the chlorosome to the reaction center may be localized.
It is the entanglement between these sites during the propagation of excitation that is discussed in~\cite{Caruso2009,Sarovar2010,Ishizaki2010,Fassioli2010,Caruso2010,Bradler2010,Scholak2011}.
(Alternatively, so called ``mode''-entanglement has also been considered in~\cite{Caruso2010}. There, the physical systems between which entanglement is investigated are no longer identical to the pigment cofactors but effective systems that are defined by the single excitation spectrum of the Hamiltonian of the coupled pigments.)

Regarding the existence and generation of entanglement in light-harvesting complexes, we draw some intuition from the following formal analogy.
As we will detail later, up to certain extent, the FMO complex can be seen analogous to a multi-armed interferometer, where each interferometer arm corresponds to a site in the FMO complex.
The propagation of a single excitation through the FMO complex is then analogous to the propagation of a single photon through the interferometer.
A single photon propagating through an interferometer can indeed immediately lead to entanglement between the arms~\cite{Enk2005}, hence from this point of view it is not very surprising that a single excitation propagating through the FMO complex may lead to entanglement between sites.
However, and this is one of our main concerns, if instead of a single photon we send a coherent state through the interferometer, then no arm entanglement will appear at all.
Indeed, at a beam splitter a coherent state is split into a product of coherent states in each of the outgoing arms.
As the light passes the various beam splitters in the interferometer, in each step we maintain a direct product of coherent states in each of the arms.
We stress that this is true even for very weak coherent states where the probability of having more than a single photon is overwhelmingly small.

There are two lessons to be learned from this analogy.
First, whether or not entanglement between interferometer arms exists depends crucially on the initial state.
Our concern is that in the case of photosynthesis in which the entire light-harvesting complex is illuminated by weak classical light and not single photons, entanglement may therefore not appear inside the FMO complex.

The second lesson is that the actual dynamics of light propagation through an interferometer, and hence a measure of the transport efficiency, does not really depend on the fact that at single photon level entanglement between arms is produced.
Indeed, the dynamics of an interferometer can very well be described at classical level, i.e.\ via coherent states, where the question of entanglement does not appear.
Hence we conclude that entanglement beyond the mere existence of coherences, although it does appear at single photon level, is irrelevant for the light transport in interferometers.

However, excitation transport in the FMO complex is not exactly identical to light propagation in an interferometer.
It is therefore important to see whether or not the problems we mentioned above are still relevant for excitation propagation through light-harvesting complexes.
In this paper we argue that, indeed, entanglement plays no essential role in transport through the FMO complex.

Following from the discussion above, we thus need to revisit the general assumptions that have been made in existing theoretical treatments of entanglement in light-harvesting complexes, in particular the following points:
\begin{enumerate}
\item the initial state entering the complex,
\item the transport between sites and the possible entanglement generated thereby.
\end{enumerate}
More specifically, we also reconsider the modeling of each of the sites as a simplified two-level system.

\section{Initial excitation}

The available literature on entanglement generated during excitation transport in pigment protein complexes usually assumes an initial state where (a) \emph{only one} site is excited and (b) this site is excited with exactly one quantum of electronic excitation, i.e.\ a ``Fock state'', as done in refs.~\cite{Caruso2009,Sarovar2010,Fassioli2010}, for example.
Given these assumptions, the pigments of the light-harvesting complex have been modeled by two-level systems, corresponding to presence or absence of an excitation at that location.

The rationale behind the above assumption for the initial state is often the following. Under illumination by sun light or, in experiment, by femtosecond laser pulses, the photon flux per pigment is weak. An estimation for conditions in full sunlight yields a value of about 10 absorbed photons per chlorophyll molecule per second in the relevant part of the spectrum~\cite{Blankenship}. Considering that the typical timescales for exciton transport is of the order of picoseconds with exciton lifetimes of nanoseconds, this suggests that \emph{if} light is absorbed, then most likely only one excitation is present.

However, our main point is that the light that excites the light-harvesting apparatus is essentially classical, technically a mixture of coherent states. Importantly even if a coherent state is very weak, and the probability of containing more than one excitation is very low, this by no means implies that it can be described as a single excitation Fock state. Hence, using the weakness of the incident light as an argument for modeling the initial state as a Fock state is a fallacy.

A coherent state differs from a Fock state in two important aspects.
First of all, a coherent state contains not only one excitation but also terms with more excitations. Second, the coherent state also contains a term, the vacuum, without excitations. In particular, the existence of the vacuum term has interesting consequences. It is completely irrelevant as far as the dynamics of excitation transport is concerned, it just represents the fact that nothing happens, but it has crucial implications to entanglement.
Since the entanglement that is considered in light-harvesting systems stems from superpositions of states in which one pigment is excited and all others are in the ground state, a contribution with all pigments being in the ground state, corresponding to the vacuum in the incoming light, plays an essential role and must not be neglected.

At this point another possibility arises, namely, that the initial state in the \emph{FMO complex} that is of interest for us, could be a Fock state for other reasons than the weakness of the incident light. Indeed, the FMO complex in vivo is not directly excited by the incident light, rather it receives its input through the antenna complex. It it thus conceivable that some mechanism in the antenna and its connection to the FMO complex may somehow generate a Fock state. In quantum mechanical terms this amounts to a state preparation process. What would this require?

Producing a Fock state in the FMO complex, would require two things. On one hand, cutting the possibility of more than one excitation. Such mechanisms have been discussed~\cite{Bruggemann2004}, although it is not clear to us how efficient they are in limiting the number of higher excitations within the time frame that is relevant for transport through light-harvesting complexes. On the other hand, to produce a Fock state, one should also be able to eliminate the vacuum component of the state. To our knowledge, no such mechanism has been suggested so far.

This entire discussion above his highly relevant since, as far as entanglement is concerned, the difference between coherent states, even if they are extremely weak, and one-photon Fock states is dramatic.
While one-photon Fock states may result in entanglement between linearly coupled systems, coherent states will not.
Furthermore, even in non-linearly coupled systems, weak coherent states can only produce very limit amounts of entanglement, far below that produced by one-photon Fock states.

\section{Harmonic oscillator model}

Having discussed the difference between the initial states, we now explicitly illustrate the difference that the two states make regarding entanglement.

In existing studies of entanglement in light-harvesting complexes, the manifold of electronic states of each of the relevant pigments, e.g.\ chlorophyll molecules, is usually restricted to only a few levels, mostly only the electronic ground state and the first excited state.
The rich energetic landscape of each chlorophyll molecule is thereby conceptually replaced by a two-level atom.

In order to account for small contributions of higher excited states, we must consider a different model for the electronic level structure of each pigment molecule, that allows for more excitations to be present at a single site.
As the simplest possible scenario, consider the analogy of a light-harvesting complex to an interferometer.
In this analogy, the electronic level structure of each site is modeled by a harmonic oscillator and thus it is formally identical to a light mode in an interferometer.

For capturing the principles of entanglement generation during the evolution of the excitations in a network of coupled chlorophyll pigments, it is sufficient to first study the interaction and state evolution of the simplest network of only two sites, a dimer. Hence, we first consider how entanglement is generated between two coupled harmonic oscillators.
(Incidentally, the present approach to model the pigment molecules by harmonic oscillators has also been employed in~\cite{Eisfeld2011}, however, by means of \emph{classical} harmonic oscillators and with a different aim, namely to illustrate that the phenomenology of quantum coherent exciton transport can also be obtained with a classical coherent model.
In the present work, however, we employ a full quantum description, and our focus lies on entanglement.)

We assume for the interaction Hamiltonian a standard form where excitations are exchanged between the modes, the rotating-wave approximation has already been applied, and the systems are taken to be resonant:
\beq \label{eq:Hamiltonian}
H_\text{int}=g \hbar (a^\dag b + a b^\dag).
\eeq
The coupling strength is denoted by $g$, and the creation and annihilation operators $a$, $b$, and $a^\dag$, $b^\dag$, for the harmonic oscillators describing molecules A and B, respectively, realize the exchange of an excitation between the molecules.

To exemplify our argument with coherent states, we investigate the initial state of a coherent state for molecule~A, which expanded into the Fock basis of states with $n$ excitations reads
\beq
\ket{\psi_A(0)}=\ket{\alpha}=\re^{-\abs{\alpha}^2/2} \sum_{n=0}^\infty \frac{\alpha^n}{\sqrt{n!}} \ket{n},
\eeq
and whose mean number of excitations is given by $\abs{\alpha}^2$.
For molecule B we take the vacuum (ground) state $\ket{\psi_B(0)}=\ket{0}$.
The requirement that the light intensity is low and hence an excitation occurs only with little probability, and that higher excited states should virtually not occur, thus formally amounts to $\abs{\alpha}\ll 1$ for the initial state. This initial state is in accordance with the central assumption that \emph{if} an excitation occurs, then most probably there is only a single excitation, because for $n>1$
\beq
\abs{\braket{1}{\psi_A(0)}}^2=\re^{-\abs{\alpha}^2}\abs{\alpha}^2 \gg \re^{-\abs{\alpha}^2} \abs{\alpha}^{2n}/n! =\abs{\braket{n}{\psi_A(0)}}^2.
\eeq
In the interferometer analogy this initial state amounts to a coherent state and a vacuum state for the two incident light modes, respectively.

Under the given Hamiltonian, the state of the two molecules evolves in the interaction picture according to
\beq \label{eq:timeEvolMol}
\ket{\psi_{AB}(t)}=U(t)\ket{\psi_A(0)}\ket{\psi_B(0)}=\re^{-igt(a b^\dag + a^\dag b)}\ket{\alpha}\ket{0}.
\eeq
The time evolution of this system of two coupled harmonic oscillators is an elementary problem.
A standard textbook calculation yields~\cite{MandelWolf}
\begin{align}
\ket{\psi(t)}
&=U(t) D_\text{A}(\alpha) U^\dag(t) U(t) \ket{0}\ket{0}\\
&=\re^{\alpha Ua^\dag U^\dag -\alpha^*UaU^\dag} \ket{0}\ket{0}.
\end{align}
Here, the displacement operator $D_\text{A}(\alpha)=\exp(\alpha a^\dag-\alpha^* a)$ is used to construct coherent states by simply moving the ground state (vacuum) away from the phase space origin, $\ket{\alpha}=D_\text{A}(\alpha)\ket{0}$. The unitarity of $U(t)$ allows to sandwich all operators in the exponential by inserting identity operations, the time argument has been omitted, and the ground state is invariant under $U$.
Using the relation,
\beq
\re^X Y \re^{-X}=Y + [X,Y] + \frac{1}{2!}[X,[X,Y]] + \frac{1}{3!}[X,[X,[X,Y]]] + \dotsb,
\eeq
the sandwiched creation and annihilation operators can be evaluated:
\begin{align}\label{conjField1}
Ua^\dag U^\dag &= \cos(gt) a^\dag +i \sin(gt) b^\dag \\
UaU^\dag &= \cos (gt) a -i \sin (gt) b. \label{conjField2}
\end{align}
Since the operators for molecule A and B commute, the exponential can be split in two parts that each give a displacement operator for molecule A and B, respectively,
\begin{align}
\ket{\psi(t)} &= D_\text{A}[\alpha\cos (gt)] \; D_\text{B}[i\alpha\sin (gt)] \; \ket{0}\ket{0} \nonumber \\
&= \ket{\alpha\cos (gt)} \; \ket{i\alpha\sin (gt)}. \label{prodCoherent}
\end{align}
In the final state after some interaction time~$t$, each of the two molecules is in a coherent state with parameters $\alpha\cos(gt)$ and $i\alpha\sin(gt)$, respectively.
With respect to entanglement, let us point out that for all interaction times and irrespective of $\alpha$, the two molecules are always in a product state, i.e.\ there is strictly no entanglement between the molecules.

\bigskip

Let us spell out the formal analogy between the dynamics in our simplified model of the two interacting chlorophyll molecules, the electronic structure of each of which is modeled by a harmonic oscillator, and two light modes interacting at a beam splitter.
At a beam splitter, in Schr\"{o}dinger picture, the state of the two incident and outgoing light modes $\ket{\phi_\text{in}}$ and $\ket{\phi_\text{out}}$, respectively, is related by a unitary operation~\cite{Yurke1986}, which contains the reflectivity/transmissivity of the beam splitter as a parameter~$\theta$:
\beq
\ket{\phi_\text{out}} = U(\theta) \ket{\phi_\text{in}} = \re^{-i\frac{\theta}{2}(a^\dag b + ab^\dag)} \ket{\phi_\text{in}}.
\eeq
The unitary operation of the beam splitter is formally identical to that of the two interacting molecules in~\eqref{eq:timeEvolMol}.
Therefore, the state of the molecules before and after they have interacted for some time $t$ can be associated to the state of the two light modes before and after they have passed the beam splitter with reflectivity/transmissivity parameter $\theta/2=gt$.
A 50/50 beam splitter with $\theta=\pi/2$ thus gives states in the output ports that correspond to the states of the molecules after time $gt=\pi/4$:
For an initial coherent state, one obtains a product of coherent states as in~\eqref{prodCoherent}.
In contrast, for a single photon entering the beam splitter in one input port, $\ket{\phi_\text{in}}=\ket{10}$, the state emerging from the beam splitter is, $\ket{\phi_\text{out}}=\big(\ket{10}+i\ket{01}\big)/\sqrt{2}$, a maximally entangled state of the two interferometer arms.

\bigskip

Let us translate the observation of a coherent state versus a single photon Fock state from the interferometer analogy back to the model system of interacting pigment molecules.
Given that the initial state is a coherent state, the act of limiting the focus of the treatment to the single excitation manifold creates the \textit{illusion} of entanglement.
Formally, one projects the total state \eqref{prodCoherent} to the subspace with exactly a single excitation and obtains for the initial state
\beq
P_1 \ket{\alpha}\ket{0} \propto \ket{10},
\eeq
where
\begin{equation}
P_1 = \proj{10} + \proj{01}
\end{equation}
is the projection onto the single excitation subspace. That is, there is one excitation on molecule~A and none on~B.
For the time-evolved state this projection yields
\begin{equation}
P_1 \ket{\psi_{AB}(t)} = P_1 \ket{\alpha\cos(gt)}\ket{i\alpha\sin(gt)} \propto \cos(gt)\ket{10}+i\sin(gt)\ket{01}.
\end{equation}
In particular for $gt=\pi/4$ the state appears to be maximally entangled in analogy to a single photon traversing a 50/50 beam splitter.

\bigskip

It is interesting to ask why does considering all possible excitations remove the entanglement? How can the higher excitations sector, which has a very small contribution, remove the entanglement? This seems to be quite paradoxical. The answer is that if the coherent state is so weak as to have only a very small amount of higher excitations, it necessarily also has a very large vacuum component. Just considering the vacuum term in addition to the single excitation sector reduces the amount of entanglement considerably. Any residual entanglement is eliminated by the higher excitation sector.

To see the effects from above, it is convenient to use a measure for the amount of entanglement as given by the concurrence~\cite{Wootters}, which for a pure state of two systems is defined in terms of the purity of one (any) of the subsystems:
\beq
C\big(\ket{\psi_{AB}}\big) = \sqrt{2\left(1-\Tr\rho_A^2\right)}
\qquad \text{with} \qquad
\rho_A = \Tr_B \proj{\psi_{AB}}.
\eeq
The concurrence of the (renormalized) state after projecting out the single excitation sector gives
\beq
C\left(\frac{P_1\ket{\psi_{AB}}}{\norm{P_1\ket{\psi_{AB}}}}\right) = \abs{\sin(2gt)},
\eeq
which is maximal for $gt=\pi/4$.

Alternatively, if the attention is not strictly limited to the single excitation manifold but the vacuum (ground state) term is considered in addition, the relevant projection to apply to the state is
\begin{equation}
P_{0,1} = \proj{00} + P_1.
\end{equation}
For the initial state, this projection yields a superposition of the ground and excited state of the first site, i.e.
\begin{equation}
P_{0,1} \ket{\psi_{AB}(0)} = P_{0,1} \ket{\alpha}\ket{0} \propto \Big(\ket{0}+\alpha\ket{1}\Big)\ket{0},
\end{equation}
and, for the time-evolved state, a superposition of the ground state and the evolved state in the single excitation manifold:
\begin{equation} \label{eq:stateP01t}
P_{0,1} \ket{\psi_{AB}(t)}
\propto \ket{00}+ \alpha\Big(\cos(gt)\ket{10}+i\sin(gt)\ket{01}\Big).
\end{equation}
The entanglement in terms of concurrence of the normalized state is therefore
\begin{equation} \label{eq:concP10t}
C\left( \frac{P_{0,1}\ket{\psi_{AB}}}{\norm{P_{0,1}\ket{\psi_{AB}}}}\right) = \frac{\abs{\alpha}^2}{1+\abs{\alpha}^2} \abs{\sin(2gt)}.
\end{equation}
We find that the maximal amount of entanglement at $gt=\pi/4$ is limited by the dominant ground state contribution for small $\abs{\alpha}$, because the ground state is not entangled.

The entanglement that appears after projecting the state onto its zero and single-excitation subspace with $P_{0,1}$ is simply rescaled with respect to the entanglement found when projecting only to the single excitation manifold.
The scaling factor $\abs{\alpha}^2/(1+\abs{\alpha}^2)$ amounts to the single excitation fraction of the projected state, i.e.\ the probability of measuring exactly a single excitation in the state once it has been projected.
As long as the assumption holds that the probability of more than one excitation is small, we require that also $\abs{\alpha}$ is very small, and therefore the entanglement reduction due to ground state contribution is larger. The higher excitation terms in the coherent state may be small in absolute magnitude, but they only have to eliminate this weak residual entanglement.

\bigskip

Another aspect that needs to be taken into account when advancing to more realistic models for the study of quantum entanglement in light-harvesting complexes is that superpositions of electronic states on the same molecule may suffer from \emph{decoherence}, primarily due to the coupling of the electronic structure to nuclear degrees of freedom.
This process ultimately turns a coherent state
\beq
\rho_A(0) = \proj{\psi_A(0)} = \proj{\alpha},
\eeq
when written as a density operator, into the incoherent mixture
\beq
\rho_A(t\to\infty) = \re^{-\abs{\alpha}^2}\sum_{n=0}^\infty \frac{\abs{\alpha}^{2n}}{n!} \proj{n}
\eeq
of electronic states of the individual molecule, meaning that after complete decoherence all coherences between states of different excitation number will have decayed.
At room temperature molecules have been demonstrated to remain in a coherent superposition of their electronic ground and first excited state for a 50\,fs-timescale~\cite{Hildner2011}, which might, however, differ for pigment molecules embedded in a protein matrix.

In the worst case, a complete decoherence of an initially coherent state and subsequent projection to ground and single excited manifold yields the incoherent mixture of the ground state and the coherently propagated single excitation:
\begin{align}
\rho_{0,1}^{(\text{decoh})} (t) &=  \frac{1}{1+\abs{\alpha}^2} \bigg[ \proj{00} \\
&+ \abs{\alpha}^2 \Big(\cos(gt)\ket{10}+i\sin(gt)\ket{01}\Big) \Big(\cos(gt)\bra{10}-i\sin(gt)\bra{01}\Big) \bigg]. \nonumber
\end{align}
The concurrence for this mixed density matrix can be evaluated with Wootter's formula~\cite{Wootters}, and gives
\beq
C \left(\rho_{0,1}^{(\text{decoh})} (t) \right) = \frac{\abs{\alpha}^2}{1+\abs{\alpha}^2} \abs{\sin(2gt)},
\eeq
that is, the same result as for the projected coherent state~\eqref{eq:concP10t}.

\section{N-level system model}

After we have formally established that the molecules are not entangled if the initial state is a coherent state, and that a projection of the state to the single excitation manifold creates the illusion that entanglement were present, let us now investigate how the number of considered levels in addition to the ground state and the single excited state affects the apparent amount of entanglement. In other words, we generalize from the projections $P_1$ and $P_{0,1}$ of the previous section to projections that include higher number of excitations.

An alternative way to look at this question is the following consideration. The electronic eigenstates of molecules do not form an infinite uniform level structure as a harmonic oscillator. It is reasonable to assume that only finitely many excitations can be supported by each pigment molecule, and any number of excitations beyond a certain threshold would cause ionization processes that take this fraction of the state out of the considered events.
For example, instead of a coherent state at one of the molecules, which involves contributions of arbitrarily many excitations,  the molecule may at most support two excitations. The initial state would thus be the first three terms of the coherent state expansion until $n=2$:
\beq \label{eq:init3}
\ket{\psi_A^{(3)}(0)} = \frac{1}{1+\abs{\alpha}^2 + \abs{\alpha}^4/2!} \left( \ket{0} + \alpha \ket{1} + \frac{\alpha^2}{\sqrt{2!}} \ket{2} \right),
\eeq
which corresponds to the projection $P_{0,1,2}$ applied to the coherent state $\ket{\alpha}$, and the result renormalized.
For these kinds of initial states, which constitute a modification of coherent states towards a more realistic initial state for pigment molecules, we now study how the number of additionally considered contributions of higher lying excited states affects the amount of entanglement that is generated during the transfer of excitations between two pigment molecules. Clearly, for the highest considered excited state being $N=1$, we recover the result~\eqref{eq:concP10t}, whereas in the limit $N\to\infty$ we recover the case of coherent states that never generate entanglement.

In general, we model each molecule by system of $N$ levels, i.e.\ the ground state and $N-1$ excited states.
In parallel with the harmonic oscillator, we choose the initial state to be the ground state for molecule~B, and a state for molecule~A,
\beq
\ket{\psi_A^{(N)}(0)}=\ket{\alpha_N}= \frac{1}{\sqrt{\mathcal{N}_{\alpha,N}}} \sum_{n=0}^{N-1} \frac{\alpha^n}{\sqrt{n!}} \ket{n},
\eeq
which is the projection of a coherent state to the lowest lying $N$ levels.
The squared norm $\mathcal{N}_{\alpha,N} =\sum_{n=0}^{N-1} \abs{\alpha}^2/n!$ approaches the value $\re^{\abs{\alpha}^2}$ found for coherent states in the limit $N\to\infty$.

As another exemplary case, let us give the analytic expressions for $N=3$ for the initial state~\eqref{eq:init3} as it evolves according to the interaction Hamiltonian~\eqref{eq:Hamiltonian}.
The time-evolved state necessarily only contains terms with at most two excitations:
\begin{multline}
\ket{\psi_{AB}^{(3)}(t)} = \frac{1}{\sqrt{1+\abs{\alpha}^2 +\abs{\alpha}^4/2}}
\bigg(
\ket{00} + \alpha \Big( \cos(gt) \ket{10} -i\sin(gt)\ket{01} \Big) \\
+ \frac{\alpha^2}{\sqrt{2}} \left( \cos^2(gt) \ket{20}  -  \sin^2(gt) \ket{02}
-i\sqrt{2}\cos(gt)\sin(gt) \ket{11} \right)
\bigg).
\end{multline}
For $\abs{\alpha}\ll 1$, the doubly excited states give only a perturbative correction to the expression of the quantum state to the case $N=2$ in~\eqref{eq:stateP01t}.
The expression for entanglement, however, does not only change by a perturbative correction, but it changes considerably.
The concurrence of the state is given by
\beq
C\left(\ket{\psi_{AB}^{(3)}}\right)
=
\frac{ \abs{\alpha}^3 \Abs{\sin(2gt)} \sqrt{8+\frac{1}{2}\abs{\alpha}^2 \big( 13+3\cos(4gt)\big)} }{ 4\left(1+\abs{\alpha}^2+\abs{\alpha}^4/2\right) }
\eeq
with its maximum at $gt=\pi/4$ of
\beq
C^{(3)}_\text{max} = \frac{ \abs{\alpha}^3 \sqrt{8+5\abs{\alpha}^2} }{ 4\left(1+\abs{\alpha}^2+\abs{\alpha}^4/2\right) }.
\eeq
The expression of the concurrence for $N=3$ is similar in structure to that of the case $N=2$ in~\eqref{eq:concP10t}, with a $\abs{\sin(2gt)}$ modulation in time, and the squared norm of the projected coherent state in the denominator.
The time-independent prefactor, however, can no longer be interpreted anymore as the probability of having an excitation in the system, as done  for $C^{(2)}_\text{max}$ in~\eqref{eq:concP10t}.
Instead, we find a scaling with the third power of $\abs{\alpha}$.

\begin{figure}
\centering
\includegraphics{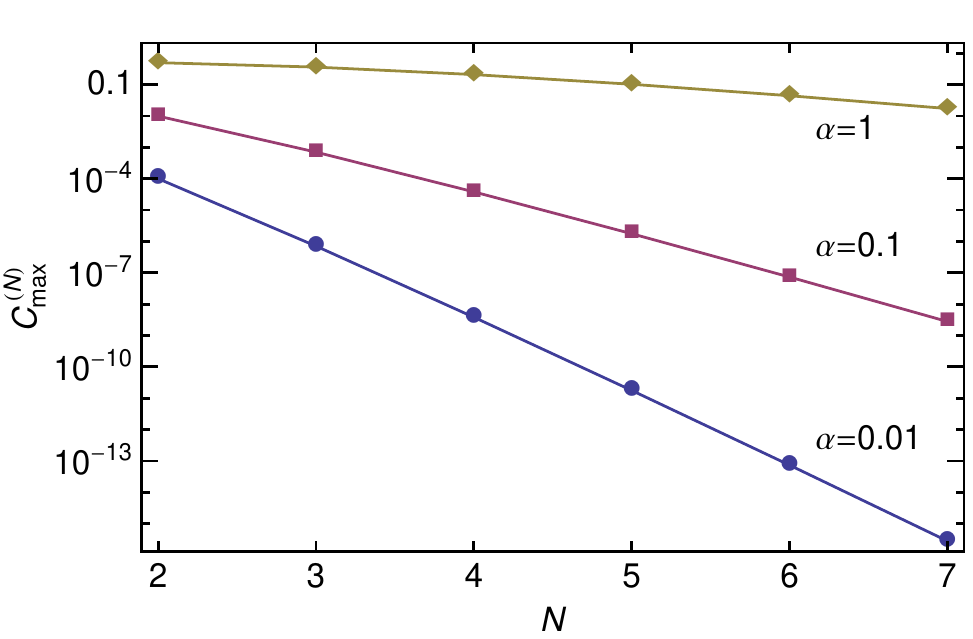}
\caption{Scaling of maximal value of apparent entanglement as measured by concurrence, $C^{(N)}_\text{max}$, for different $\alpha$ as a function of the number of considered levels $N$.
}
\label{fig:CN}
\end{figure}

Expressions for larger $N$ can be straightforwardly obtained, but are omitted here.
We have analytically examined expressions for the concurrence up to $N=7$ and generally find a global $\abs{\sin(2gt)}$ modulation in time in accordance with the intuition that maximal entanglement for these initial states is reached for the 50/50 beam splitter configuration, that is, after half the time that an excitation needs to fully move from one pigment to the other.
In particular, the expressions for maximal entanglement are of the generic form
\beq \label{eq:CN}
C^{(N)}_\text{max} = \frac{ \abs{\alpha}^N }{ \mathcal{N}_{\alpha,N} } f_N \left(\abs{\alpha}^2\right),
\eeq
that is, the apparent amount of entanglement decreases exponentially with the number of considered levels per molecule for $\abs{\alpha}<1$.
The factor $f_N$ for higher $N$ is of a similar square root form as found for $N=3$.
For $\abs{\alpha}<1$, the leading term of $f_N \left(\abs{\alpha}^2\right)$ is constant but also decreases with $N$, as listed in table~\ref{tab:fN}.
Figure~\ref{fig:CN} shows how the maximal apparent amount of entanglement decreases with $N$ for various choices of $\alpha$.

Obviously, the interpretation that given an initial coherent state for $\abs{\alpha}\ll 1$ one arrives at the single photon level is misleading because the entanglement properties of a single photon Fock state are not obtained in this limit.
Instead, for smaller values of $\abs{\alpha}$, the apparent entanglement will vanish more rapidly when increasingly many levels of the system are taken into account although they contribute to an ever smaller extent (see figure~\ref{fig:CN}).

\begin{table}
\centering
\begin{tabular}{r|ccccccc}
$N$         & 2 & 3 & 4 & 5 & 6 & 7 \\ \hline
$f_N\approx$    & 1 & $\frac{1}{\sqrt{2}}$ & $\frac{1}{4}\sqrt{\frac{7}{3}}$ & $\frac{1}{4\sqrt{2}}$ & $\frac{1}{24}\sqrt{\frac{31}{10}}$ & $\frac{1}{16\sqrt{5}}$%
\rule{0pt}{3.5ex} \\ 
$\approx$ & 1 & 0.7071 & 0.3819 & 0.1768 & 0.0734 & 0.0280
\rule{0pt}{3.5ex} \\ 
\end{tabular}
\caption{Constant leading order terms of $f_N\left(\abs{\alpha}^2\right)$ for $\abs{\alpha}\ll 1$.}
\label{tab:fN}
\end{table}

\bigskip

We have also investigated an alternative way of modeling the initial state of an $N$-level molecule by means of \emph{atomic} coherent states (also called \emph{spin} coherent states)~\cite{Zhang1990,MandelWolf}, where instead of the raising and lowering operators of the harmonic oscillator, those of the angular momentum algebra are employed.
The $N$ levels of each molecule are thereby modeled by the $N=2s+1$ levels of an effective spin-$s$ particle.
Although the computations are more intricate due to different commutation relations, we arrive at qualitatively similar results as presented for the projected harmonic oscillator.
We therefore conclude that our argument concerning the entanglement content of system that starts in a coherent state is robust with respect to the specific framework applied.

\section{Transport efficiency versus entanglement}

The \emph{excitation transfer efficiency} in chromophore complexes captures how well an excitation that starts somewhere localized in the complex traverses the network of coupled chromophores to a different location, where it is assumed to leave the complex, e.g. to the reaction center.
A question of current interest that has been addressed~\cite{Fassioli2010,Scholak2011} is, whether or not entanglement (rather than mere coherence) impacts the transport efficiency.

Let us first recollect a few essential facts about the recent treatments of excitation propagation and evaluation of transport efficiency in conjunction with the study of entanglement in light-harvesting complexes.
In a complex of coupled chromophores with pairwise interaction Hamiltonians of the kind as in~\eqref{eq:Hamiltonian}, the number of excitation quanta are conserved.
Therefore, subspaces of a fixed number of excitation quanta, e.g.\ the single-excitation subspace, evolve independently from each other.
Cross-contribution of subspaces with a fixed excitation number occur only due to interaction of the chromophores with other degrees of freedom, i.e.\ an environment, and for the excitation energies considered here ($\sim$1\,eV) mostly downward to lower numbers of excitation, due to relaxation for example.
Given that the assumption holds that the light intensity is weak and therefore the presence of higher number of excitations happens only with minuscule probability, the single-excitation subspace is essentially only subject to the unitary dynamics according to the system Hamiltonian, excitation-number conserving environment influences such as decoherence, and (excitation-number non-conserving) relaxation to the ground state.
Therefore, the single excitation-subspace evolves largely independent for the other subspaces even in the presence of decay mechanisms, and in particular it does not significantly gain excitations from higher lying states during the transfer through the complex.
Excitation transfer can only occur via the excited states, since only then an excitation (at least one) is present.
The transfer efficiency is usually evaluated by an observable, which is therefore defined only in the excited state manifold, and  for the scenarios considered here, gains its dominant contribution from the transport dynamics of the single-excitation manifold.
Contributions from higher excited states constitute only a perturbative corrections to the quantum state and therefore also to the observable that quantifies the transport properties.
Common examples for measures of the transport efficiency are obtained by integrating the population of a certain exit site, e.g. decay to the reaction center, as done in~\cite{Caruso2009}, or by the highest population of an exit-site during a certain time-window as in~\cite{Scholak2011}, for example.

Formally, for an observable $T$ that measures the excitation transport efficiency in the described way, and for initial states with only a small or even vanishing fraction of higher excited states, one has
\beq
\mean{T}\approx \mean{P_1 T P_1},
\eeq
that is, one can restrict the evaluation of $T$ with good agreement to the predominant contribution from the single excitation manifold, because higher excited states yield only a perturbative correction to this result.
Since the Hamiltonian conserves the number of excitation quanta, it is a valid approach to restrict the propagation of the entire excitation dynamics to the single-excitation subspace right from the beginning, as used in the last step of the following transformation:
\begin{align}
\mean{T}
&\approx \bra{\psi(t)}P_1 T P_1 \ket{\psi(t)} \\
& = \bra{\psi(0)} U^\dag(t) P_1 T P_1 U(t) \ket{\psi(0)} \\
& = \bra{\psi(0)} P_1 U^\dag(t) T U(t) P_1 \ket{\psi(0)}.
\end{align}
Because the projection to the single-excitation manifold is effectively carried out when measuring the transport efficiency, the projection may as well be exchanged with the dynamics to the beginning of the process.

Even with respect to open system dynamics, which is captured by the dynamical map $\Lambda(t)$, that is, the map that contains the formal solution to the Liouville equation $\dot{\rho}(t)=\mathcal{L}\rho(t)$, one can approximate the transport efficiency by only considering the single excitation manifold and the ground state:
\beq
\mean{T}=\Tr\left[T\rho(t)\right]=\Tr\left[T\Lambda(t)\rho(0)\right] \approx \Tr\left[T \Lambda(t) P_{0,1} \rho(0) P_{0,1} \right].
\eeq
Regarding the evaluation of a measure of transport efficiency it is thus, for the given assumption of weak light intensity, a natural and justified assumption to restrict the investigation to the subspace with only a single excitation.

\bigskip

In contrast to observables like measures of transport efficiencies, entanglement must be a \emph{non-linear} property of quantum states, and it can thus generally be not equivalent to coherences, which can be extracted by a linear operator.
Because of these intrinsic properties of entanglement measures, it is not possible to exchange in a similar way the dynamics in the full space with a part that is projected to the single-excitation subspace.
In fact, it is the projection to a subspace of fixed excitation number, which is a global operation, that introduces the observed entanglement into the system in the first place.
In the present case of two molecules, the projection $P_1$ acts like a Bell-state measurement.

\bigskip

From the presented case we can now provide an insight about whether or not the entanglement that may be observed in manifolds of fixed excitation number is of relevance to the state evolution or the transport properties of the system.
The transport efficiency is robust under changes of the underlying models and assumptions about the initial state regarding the presence of small perturbative corrections of higher lying excited states, whereas entanglement is not. Therefore, in a pigment protein complex where an initial excitation merely evolves according to the system Hamiltonian and under a coupling to a bath, which introduces decoherence or relaxation, entanglement cannot be a quantifier of transfer efficiency.
It cannot be the cause of a large transport efficiency, nor enhance the transport efficiency.
The propagation of an excitation in a pigment protein complex is different from the case of, say, quantum information communication tasks like quantum teleportation, where entanglement can be identified as the key resource and unentangled (separable) states cannot be used or, as another example, the fact that in interacting quantum systems entanglement is required to reach the ground state.

\bigskip

To conclude, in the first place, it is not at all clear to us that entanglement exists in the FMO complex; in fact, it seems to us that it does not. Entanglement, as it is at present postulated, relies on the existence of a single excitation in the FMO complex (technically a single excitation Fock state). As we argued, this cannot be obtained by simply having very weak light impinging on the light harvesting complex, as wrongly assumed in most literature on the subject. Consequently, if not simply due to low intensity light, the only other way in which the postulated entanglement may still exist is if there would be a dynamical, active,  mechanism of state preparation. We argue that this is also extremely unlikely. Indeed, one would need a non-trivial process based on measuring the number of excitations, and opening the entrance of the FMO complex depending on the presence of a single excitation. Crucially, this process should also eliminate the large vacuum component.
Finally, and more significantly, even if entanglement exists, its role for the transport efficiency seems to us to be irrelevant, whereas the role of coherence may be important.

\section*{Acknowlegements}
The research was funded by the Austrian Science Fund (FWF): F04011, F04012,
and the European Union (NAMEQUAM).



\label{lastpage}

\end{document}